                    \documentstyle[12pt,twoside,fleqn,espcrc1]{article}

\newcommand{\AmS}{{\protect\the\textfont2
  A\kern-.1667em\lower.5ex\hbox{M}\kern-.125emS}}

\title{Thermal Pions \hfill BUTP-93/16}

\author{Ulf-G. Mei{\ss}ner\address{Institute for Theoretical Physics,
        University of Berne, \\
        Sidlerstr. 5, CH-3012 Berne, Switzerland}%
        \thanks{Heisenberg fellow. Address after Sept. 1, 1993: CRN,
                Physique Th{\'e}orique, B.P. 20 CR, F-67037 Strasbourg Cedex 2,
                France. Work supported in parts by Deutsche
                Forschungsgemeinschaft and Schweizerischer Nationalfonds.
                Plenary talk, "Quark Matter 93", Borlaenge, Sweden, June
                1993.}}

\begin{document}
% typeset front matter
\maketitle

\begin{abstract}
I discuss the absorption and dispersion of pions in hot matter. A two-loop
calculation in the framework of chiral perturbation theory is presented and
its result is compactly written in terms of the two- and three-particle forward
$\pi \pi$ scattering amplitudes. At modest temperatures, $T \le 100$ MeV, the
change in the pion mass is small and its dispersion law closely resembles the
free space one. At these temperatures, all quantities of interest are given to
a good degree of accuracy by the first term in the virial expansion which is
linear in the density.
\end{abstract}

\section{INTRODUCTION}

The properties of pions in hot hadronic matter are encoded in the pion
propagator, in particular in the mass shift due to the influence of the
heat bath,

\begin{equation}
(p^0)^2 = \vec p\,^2 + M_\pi^2 + \Pi (p^0 , \vec p \,)
\end{equation}

where $M_\pi$ denotes the pion mass, $p^0$ its energy, $\vec p$ its
three-momentum and $\Pi$ its self-energy, which is
in general a complex quantity. Its real part is related to the dispersion and
the group velocity while its imaginary part encodes the information about the
pion absorption. Chiral perturbation theory (CHPT) allows to systematically
calculate the behaviour of $\Pi$ at low and modest temperatures as will be
spelled out below. This calculation is done under the following assumptions.
I assume a hadron gas in a state of thermal equilibrium which mostly consists
of pions. Therefore, $T$ should not be larger than approximately 100 MeV
because at higher temperatures the massive states start to dominate
\cite{gerb}.
In particular, there should not be any sizeable baryon contamination in the
gas. Therefore, the results which will be presented below are mostly relevant
for future experiments and facilities which will have more pions and reach
lower temperatures than the present ones. It is instructive to briefly recall
the scattering of light on a dilute gas of $N$ molecules in a volume $V$. The
index of refraction will in general be complex and to leading order in the
density is given by \cite{gold}:

\begin{equation}
n + i \kappa \simeq 1 + \frac{2 \pi N}{k^2 V} \, f
\end{equation}

with $f$ the forward scattering amplitude of photons on a single molecule.
Making use of the optical
theorem, it follows that $\kappa$ is directly proportional to the total forward
scattering cross section. In complete analogy, we will see that the absorptive
and dispersive properties of the pions are related to the forward $\pi \pi$ and
$\pi \pi \pi$  scattering amplitudes (the latter is related to the effects of
second order in the density).

\section{EFFECTIVE FIELD THEORY OF QCD AT FINITE TEMPERATURE}

At low energies, the QCD Green functions are dominated by the (almost) massless
Goldstone bosons related to the spontaneous breakdown of the chiral symmetry,
$SU(3)_L \times SU(3)_R \to SU(3)_V$. Furthermore, to a good first
approximation the quark masses can be set to zero (which defines the
chiral limit of QCD). The consequences of these features can be worked out in a
systematic fashion by making use of an effective field theory,

\begin{equation}
{\cal L}_{\rm QCD} ={\cal L}_{\rm eff}[ U, \partial_\mu U, {\cal M}]
\end{equation}

where $U = \exp [ i\pi / F_\pi ]$ embodies the Goldstone (pion) fields and
${\cal M}$ is the quark mass matrix. A fundamental scale of the strong
interactions at low energies is set by the pion decay constant $F_\pi$, which
is defined via $<0|A_\mu^{i}|\pi^k> = i \delta^{ik} p_\mu F_\pi$, $F_\pi \simeq
93$ MeV, with $A_\mu^i$ the axial current. One can systematically expand all
Green functions around the chiral
limit. This is a simultaneous expansion in small momemta, small quark (pion)
masses and small temperatures, the corresponding small parameters being
$p/\Lambda$, ${\cal M}/\Lambda^2 $ and $T/\Lambda$. Here, $\Lambda \simeq
M_\rho$ is the scale where the non-Goldstone excitations become important. As
shown by Weinberg \cite{wein}, in the effective Lagrangian framework this
amounts to an expansion in pion loops. The effective Lagrangian consists of a
string of terms with increasing number of derivatives and/or quark mass
insertions,

\begin{equation}
{\cal L}_{\rm eff} = {\cal L}^{(2)} + {\cal L}^{(4)} + \ldots
\end{equation}

To lowest order, one calculates tree diagrams using ${\cal L}^{(2)}$. This
leads
to the venerable current algebra results. At next-to-leading order, one has to
calculate one-loop diagrams using ${\cal L}^{(2)}$ to perturbatively restore
unitarity and tree diagrams with
exactly one insertion from ${\cal L}^{(4)}$. The coefficients of the latter
terms are not restricted by symmetry but have to be determined from
phenomenology \cite{gl}.

At finite temperatures, only a few modifications occur. First, if one works in
a real-time approach, one has to choose a proper path for the time integration
in the action. This path is called $\cal C$ and extends from $-\infty$ to
$+\infty$ above the real axis, returns below the real axis and goes down to
$-\infty - i\beta$ parallel to the imaginary axis \cite{lm}. Here, $\beta = 1 /
T$. Second, the pion propagator is modified according to well-known rules and
finally, the conventional time-ordering at $T=0$ is substituted by
path-ordering along $\cal C$ for $T\ne 0$, e.g.

\begin{equation}
<0 | T ( A_\mu A_\nu ) | 0> \,\,\, \to \,\,\, <0 | T_{\cal C} ( A_\mu A_\nu ) |
0> \end{equation}

More details on CHPT and its application to finite temperature physics can be
found e.g. in the review \cite{um} or in Gerber and Leutwyler \cite{gerb}.

\section{TWO-LOOP CALCULATION OF THE PION SELF-ENERGY}

\subsection{One-loop calculation and virial expansion}

It is instructive to recall how one calculates the pion properties in CHPT at
$T=0$.
One particularly useful choice of the interpolating pion field is
the axial current due to the PCAC relation. The correlator of two axial
currents reads \cite{gl}

\begin{equation}
i \int dx \, e^{i p x}<0 | T ( A_\mu^{i}(x) A_\nu^k(0) ) | 0> = \delta^{ik}
\left\{  \frac{p_\mu p_\nu F_\pi^2}{M_\pi^2- p^2}
       + g_{\mu \nu} F_\pi^2 + \ldots \right\}
\end{equation}

This can straightforwardly be extended to the finite temperature case. Another
method is the virial expansion to which I will come back later.
The calculation of the pion self-energy to one loop order has been done by
Goity and Leutwyler \cite{goi}, Shuryak \cite{shu1} and Schenk \cite{s1}. One
splits the
pion self-energy as $\Pi = \Pi^0 + \Pi^T$, where $\Pi^T$ vanishes at $T=0$ and
describes the modification due to the heat bath. Instead of Re $\Pi$ and Im
$\Pi$, one conventionally writes

\begin{equation}
p^0 = \omega (p) - \frac{i}{2} \gamma (p)
\end{equation}

with $\omega (p)$ the frequency of the pionic waves and $M_\pi (T) = \omega
(p=0)$ the effective mass. The damping coefficient $\gamma (p)$ is the inverse
time within which the intensity of the wave is attenuated by a factor $1/e$.
In the one-loop approximation, one simply finds

\begin{equation}
\omega (p) = \sqrt{\vec p \,^2 + M_\pi^2 (T)} ,\,\,\,\,
\gamma (p) = 0 \end{equation}

This is a quite common phenomenon in CHPT calculations - to get to the
imaginary
part of a certain quantity with the same accuracy as the real part, one has
to work harder. It is simply related to the fact that to lowest order one
calculates tree diagrams, which are real. The temperature-dependent pion mass
$M_\pi (T)$ has e.g. been given by Gasser and Leutwyler \cite{glt} in the
framework of
CHPT. Notice that there is no absorption to one loop, the only effect of
the interaction with the heat bath is that the pion mass and decay constant
become $T$-dependent. There is another way of presenting the one-loop result.
If one inspects
the corresponding Feynman diagrams, one sees that one effectively deals with
the $\pi \pi$ scattering amplitude in forward direction, $T_{\pi \pi} (s)$ with
$\sqrt{s}$ the cm energy of the collision. To this order, the pion self-energy
is linear in the density and one can write

\begin{equation}
\Pi^T (\omega_p, \vec p \,) = - \int \frac{d^3 q}{(2\pi )^3 2 \omega_q} \,
n_B (\omega_q ) \, T_{\pi \pi} (s)
\end{equation}

with $\omega_k = \sqrt{\vec k \,^2 + M_\pi^2}$ and $n_B^{-1} = (\exp[\beta
\omega] - 1)$ is the canonical Boltzmann factor. This is also the first term in
the virial expansion. For a general proof, see \cite{lue}. This last
formula is very powerful. If one uses empirical input for $T_{\pi \pi} (s)$,
this form allows one to go beyond the strict one loop CHPT result. In
particular, the important effect of the $\rho$ resonance can thus be
implemented fully, whereas in the one-loop representation of $T_{\pi \pi} (s)$
one only sees the tail of the $\rho$ (for more details see e.g. \cite{s1}).
One result of employing this procedure is that
the pion mass depends only very weakly on the temperature \cite{s1}.

\subsection{Two-loop calculation}

Schenk \cite{s2} has extended the above to two loops. This calculation is much
more tedious - there are not only more diagrams, but one also has to account
for
three-particle scattering as inspection of some Feynman diagrams reveals. In
particular, one finds that

\begin{equation}
\Pi^T = \Pi^{(1)} + \Pi^{(2)}
\end{equation}

which means that one has contributions which are linear and quadratic in the
Boltzmann factor (density). The diagrams which contribute to $\Pi^T$ involve
insertions from ${\cal L}_{\rm eff}^{(2)}$ and ${\cal L}_{\rm eff}^{(4)}$,
the contribution from ${\cal L}_{\rm eff}^{(6)}$ is temperature-independent.
Therefore, the final result for $\Pi^{(1)}$ is entirely fixed in terms of
$M_\pi$, $F_\pi$ and four low-energy constants determined already in \cite{gl}.
The term of second order in the density involves an integration over three
pion momenta and a function which is of fourth order in $M_\pi$ and the pion
momenta. The result is correct up to and including order $p^6$ in CHPT. The
explicit formulae are given in \cite{s2}. Instead of presenting
the rather lengthy expressions, let me rather discuss a more compact form
involving S-matrix elements of two- and three-pion scattering. One expects that
at two-loop order the thermal distribution will depend on the effective mass,
i.e. on $n_B (\omega_p^T )$ with $( \omega_p^T )^2 = {\vec p}\, ^2 + M_\pi^2
(T)$. In addition, there are the diagrams related to three-particle scattering.
Now the forward limit of the $3 \to 3$ scattering amplitude does not exist due
to diagrams where one pion is exchanged between clusters and is close to its
mass shell. One thus defines a proper amplitude $\hat T_{33}$ \cite{s2}

\begin{equation}
\hat T_{33} = T_{33} + \sum_q \, ' \, T_{22} \frac{1}{q^2 - M_\pi^2} T_{22}
\end{equation}

which has a well-defined forward limit ($T_{22}$ is the two-particle scattering
amplitude). For a more detailed discussion see \cite{s2}. Consequently, one can
express the result of the two-loop calculation as

\begin{eqnarray}
\Pi^T (\omega_p, \vec p \,) = - \int \frac{d^3 q}{(2\pi )^3 2 \omega_q^T} \,
n_B (\omega_q^T ) \, T_{\pi \pi} (s) \nonumber \\ - \frac{1}{2} \int \frac{d^3
q}{(2\pi )^3 2 \omega_q} \, \frac{d^3 k}{(2\pi )^3 2 \omega_k} \,
n_B (\omega_q ) \, n_B (\omega_k ) \, {\hat T}^R_{\pi \pi \pi} (p, q, k )
\end{eqnarray}

with ${\hat T}^R_{\pi \pi \pi}$ the proper retarded $3 \to 3$ forward
scattering
amplitude. Notice again that in the first term on the r.h.s. of $\Pi^T$ the
effective pion mass appears, i.e. the thermal distribution of the pions
actually depends not longer on the bare mass as in the one-loop case. The
second term is the new one, it is believed to be the most general term
quadratic in the density (which is, however, not strictly proven). Notice that
one could also use $T$-dependent energies in the $n_B^2$ term, but this goes
beyond the accuracy of the two loop calculation.  Let me
stress again that on can not further simplify this result and express e.g.
in terms of a $T$-dependent two-particle  scattering amplitude only. For
details, see \cite{s2}. The damping rate $\gamma (p)$ follows using unitarity
form eq.(12). If one neglects Bose correlations in the initial and final
states, it can compactly be written as

\begin{equation}
\gamma (p)  = \omega_p^{-1} \int \frac{d^3 q}{(2\pi )^3 2 \omega_q} \,
n_B (\omega_q ) \, \sqrt{ s ( s - 4 M_\pi^2 )} \sigma_{\pi \pi} (s)
\end{equation}

Indeed, the r.h.s. of this formula represents the collision rate for pions with
momentum $p$ moving through a pionic target whose momenta are distributed
according to the Bose factor $n_B (\omega )$. The mean damping rate is defined
by

\begin{equation}
<\gamma (p)> = \int d^3 p \, n_B (\omega_p ) \gamma (p) /
\int d^3 p \, n_B (\omega_p )
\end{equation}

\subsection{Choice of the $\pi \pi$ scattering amplitude}

Before presenting results, a few remarks on the $\pi \pi$ scattering amplitude
are in order. One can rewrite eq.(9) in terms of a thermal weight function,
$W(p,s)$. An elementary consideration leads one to conclude that for
temperatures $T < 200$ MeV the exponential behaviour of $W(p,s)$ suppresses the
scattering contributions with cm energies $\sqrt{s} \le 1$ GeV. In this energy
range, only the S and P partial waves are non-negligible, so that one can write
the forward $\pi \pi$ scattering amplitude as

\begin{equation}
T_{\pi \pi} (s) = \frac{32 \pi}{3} \left( T^0_0 + 9 T^1_1 + 5 T^2_0
\right)
\end{equation}

where the upper index denotes the isospin. As it is well-known, while the
isospin zero S-wave and the P-wave are attractive, the smaller exotic S-wave is
repulsive and thus there are cancellations in $T_{\pi \pi} (s)$. As is
discussed in detail in ref.\cite{gm}, the one-loop CHPT results are reliable up
to energies of $\sqrt{s} \simeq 500$ MeV. Beyond this energy, the corrections
become too large and also, one misses completely the peak due to the $\rho$ in
$T^1_1$. Finally, unitarity is violated above 700 MeV. Therefore, to make full
use of the virial expansion, one can either use a semi-phenomenological
parametrization of the phase shifts imposing CHPT constraints at low
energies \cite{s1,s2} or use an extended effective Lagrangian including also
the low-lying resonace fields $R$, ${\cal L}_{\rm eff} [U , R]$. Such an
approach has been shown to be successful in the description of $\pi \pi$ and
$\pi K$ scattering data \cite{bkm}. In the following section, I will therefore
present results based on the exact two-loop CHPT result (which is correct to
order $p^6$) and also making use of the virial expansion and a parametrization
of the $\pi \pi$ scattering amplitude which extends to energies of about 1 GeV.

\section{RESULTS AND DISCUSSION}

Here, I will only present the most prominent features of the extensive study
presented in ref.\cite{s2}. These can be summarized as follows:

\medskip

First, let us consider the temperature dependence of the
effective mass $M_\pi
(T)$. If one uses the full CHPT two loop result, one finds that for $T = 100
(150)$ MeV the mass is lowered by 2.5 (14) percent. Also, the effects of second
order in the density are tiny, they amount to a shift of 2.5 percent at 150
MeV. These $n_B^2$ corrections stem almost entirely from three-body collisions,
the effects of the mass shift in the Bose factor are negligible. The full two
loop CHPT result is also not very different from the first term in the virial
expansion making use of a better description of the forward $\pi \pi$
scattering amplitude. Notice that the current algebra result for the $\pi \pi$
forward scattering amplitude leads to an increasing pion mass because the
momentum-dependent part cancels in eq.(15) and the remainder is constant and
negative.

\medskip

For the quasi-particle energy $\omega (p)$ one finds that the
two-loop CHPT result and the analysis based on the phenomenological description
of the $\pi \pi$ scattering amplitude agree within 20 per cent at temperatures
and momenta below 150 MeV. At higher momenta, the influence of the $\rho$
becomes more and more pronounced and the straightforward CHPT result is not
reliable any more. Most important, however, is the fact that temperature
effects on the dispersion are small. The dispersion law of pions in a medium
closely resembles the one in free space. At $T = 150$ MeV, the scattering
with the gas modifies the fequency by less than 20 per cent. This means that
in the low $T$, long-wavelength limit one does not find any indication for a
substantial change in the dispersion law as suggested by Shuryak \cite{shu2}.
Again, the effects of order $n_B^2$ are small, i.e. the first term in the
virial expansion dominates.

\medskip

Let us now consider the mean damping rate. The effects of Bose correlations are
of the order of a few per cent, indicating again that the main contribution to
the damping rate stems form the terms of first order in the density. It is
important to notice that since the damping rate is proportional to the
imaginary parts of the partial waves (which are positive due to unitarity), no
cancellations occur betwen the S- and P-waves. As a consequence, using the
imaginary parts of the one loop CHPT prediction for the $\pi \pi$  scattering
amplitude leads to
a result which agrees within 30 per cent with the virial expansion at $T < 100$
MeV. Using the full one-loop CHPT result \cite{gm}, i.e. the real and the
imaginary parts of the partial waves, the mean damping grows much too
strongly as $T$ increases. This is due to the fact of the unitarity violation
in the $I = 0$ S-wave above 600 MeV cm energy (notice that at $T = 100$ MeV,
both collision partners have energies of about 300 MeV). For temperatures
between 100 and 200 MeV, the leading term in the virial expansion predicts a
decrease of the mean damping rate by a factor of two. For these temperatures,
however, one has to account for the massive states like the $K$, $\rho$,
$\eta$, $\ldots$. These increase the damping rate. For temperatures above
100 MeV, one expects that $<\gamma (p)> \sim T^5 / 12 F_\pi^4$ \cite{beb}.

\medskip

To summarize, I have shown that the propagation properties of pions in hot
matter are determined by the pole position of various Green functions, like
e.g. $<0|T_{\cal C} (AA)|0>$. The pion energy is in general complex, $p^0 =
\omega (p) - i \gamma (p) / 2$. At low temperatures, the virial expansion of
$\omega (p)$ and $\gamma (p)$ is appropriate. To first order in the density,
the frequency $\omega (p)$ and the damping rate $\gamma (p)$ are determined by
the forward $\pi \pi$ scattering amplitude, $T_{\pi \pi} (s)$. To second order
in the density, one has additional contributions related to the proper
three-particle scattering amplitude $\hat T_{\pi \pi \pi}$ and the effective
pion mass enters the Boltzmann factor in the first term of the virial
expansion.
However, the effects of order    $n_B^2$ are small. The main contributions to
$\omega (p)$ and $\gamma (p)$ stem from the first term in the virial expansion.
Furthermore, the temperature effects on the energy and mass are small. The
dispersion law of pions at finite temperatures closely resembles the one in
the vacuum.

%\begin{figure}[htb]
%\begin{minipage}[t]{80mm}
%\framebox[79mm]{\rule[-26mm]{0mm}{52mm}}
%\caption{Good sharp prints should be used and not (distorted) photocopies.}
%\label{fig:largenenough}
%\end{minipage}
%
%\hspace{\fill}
%
%\begin{minipage}[t]{75mm}
%\framebox[74mm]{\rule[-26mm]{0mm}{52mm}}
%\caption{Remember to keep details clear and large enough to
%         withstand a 20--25\% reduction.}
%\label{fig:toosmall}
%\end{minipage}
%\end{figure}
%

\section{ACKNOWLEDGEMENTS}

I am grateful to Andreas Schenk for patient tuition on the material presented
here.

\end{document}